\documentclass[twocolumn,amsmath,amssymb,aps,prb]{revtex4}

\usepackage{graphicx}
\usepackage{dcolumn}
\usepackage{bm}

\begin{document}

\preprint{PREPRINT}

\title{Measurements of the steady streaming flow around oscillating
spheres using 3D particle tracking velocimetry}

\author{Florian Otto}
\altaffiliation[Currently at ]{Institut fur Experimentelle und Angewandte Physik, Universitat Regensburg, D-93040 Regensburg, Germany}
\author{Emmalee K. Riegler}
\author{Greg A. Voth}

\affiliation{Department of Physics, Wesleyan University, Middletown
CT 06459, U.S.A.}

\homepage{http://gvoth.web.wesleyan.edu/lab.htm}

\date{\today}

\begin{abstract}
Granular particles vibrated in a fluid have been found to exhibit
self-organization with attractive and repulsive interactions
between the particles. These interactions have been attributed to
the steady streaming flow around oscillating particles. Here we
examine the steady streaming flow surrounding a vertically
oscillating sphere using three dimensional particle tracking
velocimetry. We present measurements of the flow with the sphere
far from boundaries, close to the bottom wall of the tank, and
near another oscillating sphere. The steady velocity field is
found to disagree with available analytic calculations. When the
sphere is oscillated near the bottom wall the entire topology of
the flow changes, resulting in a larger repulsive region than
expected. Previous experiments saw attraction between particles in
the region where the flow around a single particle is repulsive.
We conclude that advection in the streaming flow due to a single
particle cannot explain the observed attractive and repulsive
interactions, rather non-linear interactions between the flows
around two or more spheres must be responsible.
\end{abstract}

\maketitle

\section{\label{sec:intro}Introduction}

Steady streaming flows due to the oscillatory motion of a fluid
are responsible for a wide range of fluid phenomena. Work on this
problem goes back to Rayleigh~\cite{rayleigh:1883}. Applications
of steady streaming flows include driving fluid with ultrasonic
beams~\cite{humphrey:1989}, microfluidic
transport~\cite{hilgenfeldt:2004}, vesicle
license~\cite{marmottant:2003}, and streaming flows in the
ear~\cite{lighthill:1992}.

Recent experiments have observed self-organization of granular
particles when they are vibrated
vertically~\cite{voth:2002,gollub:2004} or
horizontally~\cite{garrabos:2002} in a fluid. Streaming flows were
identified as the mechanism for self-organization, but there
remained puzzling questions about the structure of the streaming
flows and the attractive and repulsive interactions between
particles. In this paper we present experimental 3D particle
tracking measurements of the flow around vertically oscillating
spheres. We show that for a streaming flow far from boundaries the
available analytic solutions do not agree with our measurements,
likely because the theories are for limiting cases that are far
from the experimental parameters. Measurements of the streaming
flow at different distances from a solid bottom wall reveal that
the wall changes the topology of the streaming flow. We also
present measurements of the flow around two oscillating spheres.

\section{\label{sec:parameterspace}Parameter Space for Oscillating
Spheres}
 \begin{figure}[b]
 \begin{center}
 \includegraphics[width=3.0in]{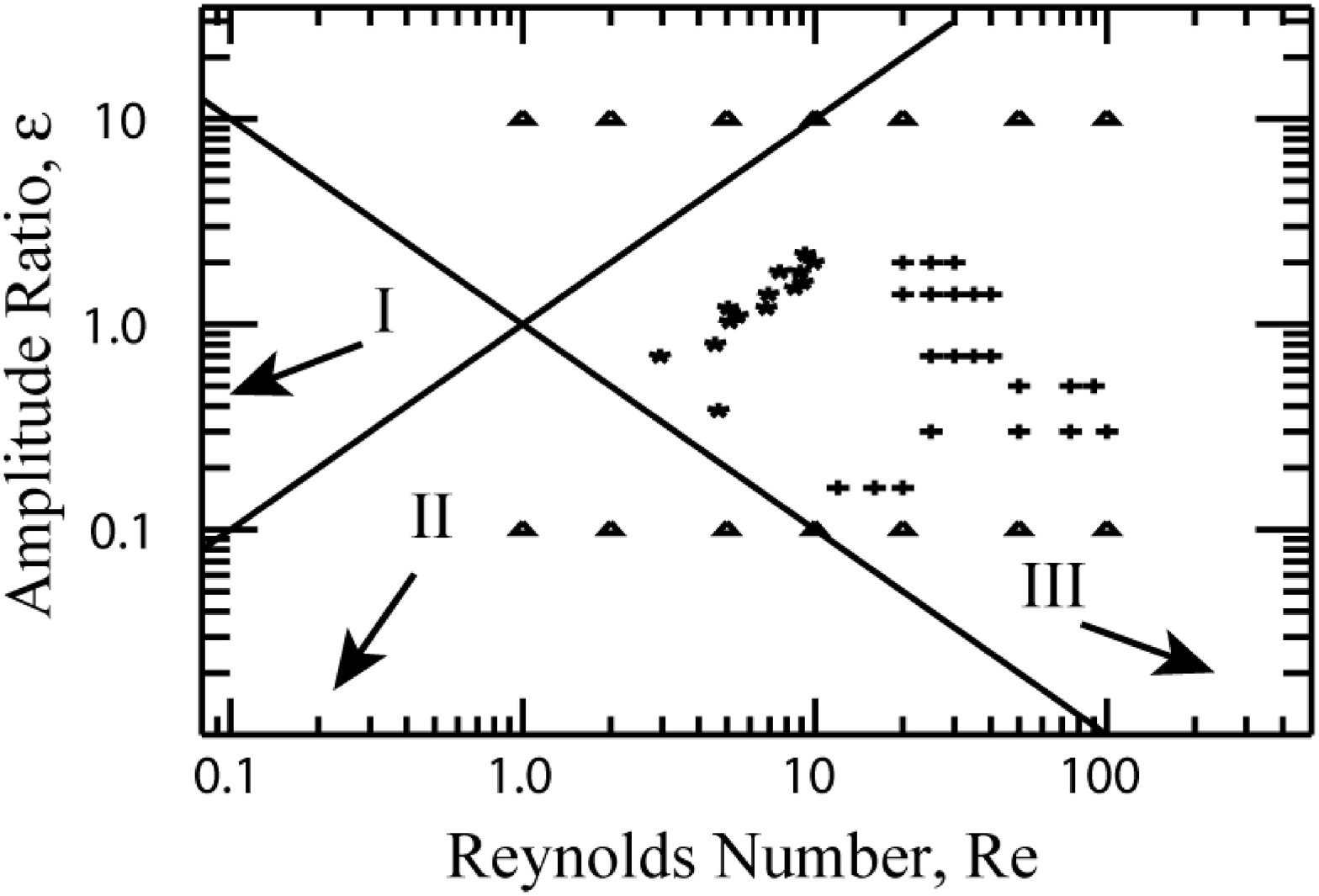}
 \caption{Parameter space for a single oscillating sphere in an
infinite fluid. Plus symbols indicate data from this study;
asterisk indicate data from Voth et. al~\cite{voth:2002};
triangles indicate data from Blackburn~\cite{blackburn:2002}. The
Roman numerals indicate the limits described in
Section~\protect{\ref{sec:parameterspace}} for which calculations
of the steady stream flow field are available.}
 \label{fig:Re-eps}
 \end{center}
 \end{figure}

In Figure~\ref{fig:Re-eps} we show the parameter space relevant for
a single oscillating sphere far from any boundaries.  The vertical
axis is the oscillation amplitude (half peak to peak)
nondimensionalized by the particle radius,
\begin{equation}
\epsilon=A/a \enspace.
\end{equation}
The horizontal axis is the Reynolds number,
\begin{equation}
Re=A\omega 2 a/\nu \enspace.
\end{equation}
Analytical solutions of the steady streaming flow around an
oscillating sphere have three limits, which are indicated by Roman
numerals in Figure~\ref{fig:Re-eps}.  The theoretical calculations
are usually described in terms of alternate nondimensional
paramters

\begin{eqnarray}
 M^2 = \frac{Re}{\epsilon} = \frac{2a^2}{\nu/\omega}
 \label{eq:MRE} \\
 \nonumber{Re_s = \epsilon \: Re= \frac{2A^2}{\nu/\omega}}
 \enspace.
\end{eqnarray}

The quantity $\delta_{osc}=\sqrt{\nu/\omega}$ is the thickness of
the oscillatory boundary layer. In Fig.~\ref{fig:Re-eps}, solid
lines are drawn at $M^2=1$ (upward sloping) and $Re_s=1$ (downward
sloping). The analytical solutions in limits I and II were found
by Riley~\cite{riley:1966}. Both these solutions are in the limit
of small $Re_s$ (so that the streaming flow is dominated by
viscosity) and small $\epsilon$ (so that the oscillation amplitude
is infinitesimal). The solution in limit I requires $M^2 \ll 1$ so
that the oscillatory boundary layer thickness is much larger than
the size of the sphere.  In this limit the streaming flow is
inward along the oscillation axis and outward at the equator.  The
solution in limit II requires $M^2 \gg 1$ so that the oscillatory
boundary layer is much smaller than the sphere size. Here the
streaming flow has an inner recirculation zone and an oppositely
directed outer recirculation zone that flows inward toward the
equator and outward along the oscillation axis (see Van
Dyke~\cite{vandyke:1982} or Fig.~\ref{fig:model-flow} for images
of this flow). Solution III was obtained by Brenner and
Stone~\cite{voth:2002}, who assume that the outer streaming flow
is potential flow ($Re_s \gg 1$ and $\epsilon \ll 1$). The flow
field here is qualitatively similar to Solution II.

The symbols in Fig.~\ref{fig:Re-eps} show the parameters of the
experiments in this paper, Voth et. al~\cite{voth:2002}, and the
simulations by Blackburn~\cite{blackburn:2002}. Despite agreement
between Solution III and experiments on spheres bouncing above a
vibrating plate~\cite{voth:2002}, we will see that the existing
theories are inadequate to quantitatively describe the streaming
flows in our experiments. Factors that contribute to the
discrepancy include the non-infinitesimal oscillation amplitude,
the intermediate Reynolds number of the streaming flow, the
effects of the bottom wall, and non-linear interactions between
the flow around two spheres.

\section{\label{sec:setup}Experimental Setup}
\begin{figure}[!tb]
\begin{center}
\includegraphics[width=3in]{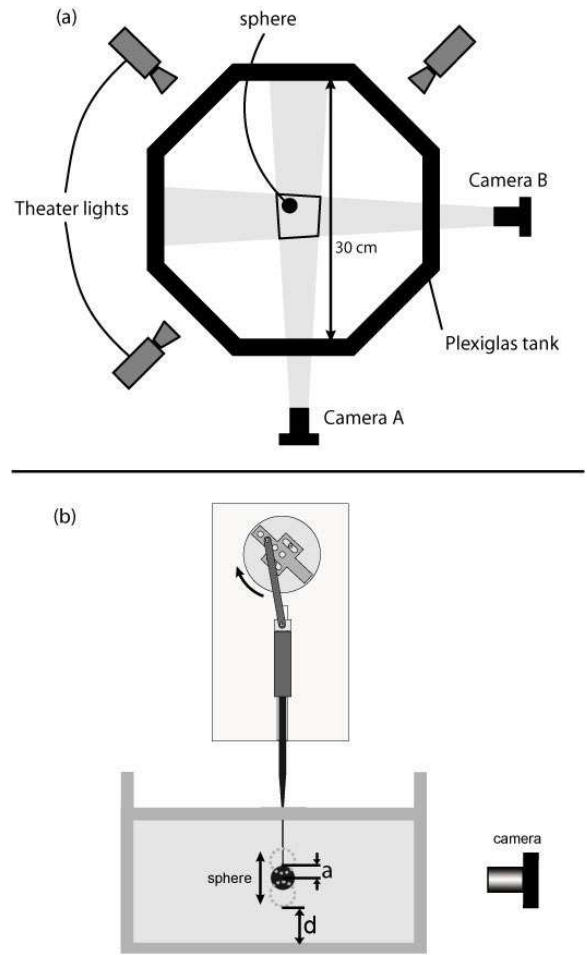}
\caption{(a) A top view of the experimental setup drawn to scale.
Two cameras at $90^\circ$ from each other provide a stereoscopic
view of a volume of the the flow around the oscillating sphere.
The overlap of the cameras is indicated by the quadrilateral.
Three theater lights provide symmetric illumination for both
cameras. (b) A side view of the setup not drawn to scale. The
circular motion of the disk sitting on a stepper-motor is
converted to a linear up-and-down motion by the use of a
connecting rod and a linear motion rail. The distance between the
lowermost point of the sphere during its stroke and the bottom
plate will be referred to as $d$.} \label{fig:tankview}
\end{center}
\end{figure}

A diagram of the experiment is shown in Figure~\ref{fig:tankview}.
The sphere is held in the center of an octagonal plexiglas tank.
The tank has an inner diameter of 30 cm and a height of 20 cm. The
octagonal shape was chosen because it nearly has cylindrical
symmetry, but still provides eight flat walls for optical access.
The sphere is attached to a thin rod that moves on a linear motion
rail.  The vertical oscillations are driven by a stepper motor as
shown in Fig.~\ref{fig:tankview}(b).  We primarily used a 1.91 cm
diameter sphere attached to a 0.16 cm diameter rod. For some
experiments we used a larger 5.09 cm sphere on a 0.33 cm rod. The
vertical position of the sphere and the amplitude of the stroke
could be adjusted continuously. The tank was filled with a mixture
of glycerol and water, adjusted to give the required viscosity.

Three-dimensional particle tracking (3DPTV) was used to
reconstruct the three dimensional positions and the velocities of
the particles in the flow from stereoscopic camera
images~\cite{dracos:1996}. 3DPTV has become a standard technique
in many areas of fluid measurement, particularly in the study of
turbulent flows~\cite{mann:2000, luthi:2005, bodenschatz:2006}. As
we show here, it can also be used to rapidly map the full three
dimensional flow field of simpler flows. Compared with PIV, 3DPTV
requires lower seeding density, so it has lower spatial
resolution. For periodic flows like ours, phase averaging can be
used to obtain high spatial resolution for the full three
dimensional flow field from a single data set. We have implemented
a simple 3DPTV setup using two 1.3 Megapixel high-speed cameras
for stereoscopic imaging. This limits the seeding density to
significantly less than is possible with 3 or 4
cameras~\cite{dracos:1996}, but the setup was sufficient for our
needs. Each camera has 4 GB of RAM which allowed 3200 images to be
acquired continuously. The cameras imaged a volume surrounding the
sphere roughly $8 \times 8 \times 8 $cm$^3$ as shown in
Fig.~\ref{fig:tankview}(a). The tracer particles were
alumino-silicate microspheres 80-120 $\mu m$ in diameter with
density 0.7-0.9 $g/$cm$^3$ (supplied by Trelleborg Fillite, Inc.).
Three 750 Watt theater lights gave symmetric illumination.

To find the three dimensional coordinates of the tracer particles,
we match the image of a particle on one camera to the image of the
same particle on the other camera. Each position on a camera image
corresponds to a particle that exists somewhere along a ray in
space. The three dimensional position of the particle is
determined by finding rays from the two cameras that intersect.
Converting image positions to rays in space requires a camera
model and an accurate calibration to determine the parameters of
the model. We used a simple distortion free camera model with
seven parameters defining camera location, viewing direction,
rotation and magnification. To obtain the calibration parameters,
we imaged a calibration mask in the glycerol mixture and measured
the real space positions of each dot. A non-linear fit of these
measured positions to the calibration model gives the necessary
calibration parameters for that camera. The system could determine
the three dimensional position of a 100 $\mu m$ particle with an
accuracy of about 10 $\mu m$.

Since we are primarily interested in the streaming flow created by
the oscillatory motion of spheres, we typically take one image per
period of the sphere oscillation. Tracking particles through these
phase locked images allows direct measurement of the steady flow.
The cameras were triggered by a photogate aligned with 16
uniformly spaced holes around the circular disk. Usually we
covered all but one hole to resolve only the streaming flow. For
some experiments we took 16 images per period to observe the
oscillatory flow, i.e. how the recirculation zones changed during
one driving cycle.

Experimental difficulties limited our access to the full parameter
space of our problem. Firstly, the rod which is attached to and
drives the sphere had some influence on the flow field. It
produced an up/down asymmetry in the streaming flow even when the
sphere was far from the wall, where up/down symmetry is usually
expected. This effect became less important as $Re_s$ was
increased. Secondly, the continuous input of heat from the theater
lights caused some convection in the fluid. We installed infrared
mirrors and absorbing glass, but where the streaming flow is very
weak (at small $Re$, small $\epsilon$, or far from the sphere) the
turbulent convective flow is still visible in the data. Finally,
at high glycerol concentrations (for many runs we used $92\%$
glycerol) we found that the index of refraction of the fluid
became large enough that polystyrene particles, which we had
planned to use as tracer particles, became invisible. The
replacement particles had a density of ($0.7-0.9 g/cm^3$) which
was significantly less than that of the fluid ($1.24 g/cm^3$), so
the particles would rise very slowly due to buoyancy. This had a
negligible effect on the measured velocities, but after a long
experimental run, there would be no tracer particles left in the
region around the sphere.   All three of these problems were less
significant for strong steaming flows at higher $Re$ and
$\epsilon$. This is why our data is at higher Reynolds numbers
than the data of Voth et. al~\cite{voth:2002} and why we did not
measure smaller $\epsilon$ to more directly compare with the
analytical results (see Fig.~\ref{fig:Re-eps}).

\section{\label{sec:results}Experimental Results}

We first looked at the flow field around a single sphere
oscillating in the center of the tank, where no boundaries should
affect the flow.  We went on to study the changes that result from
the presence of the bottom plate, where we observed a dramatic
change in the flow topology. The last set of experiments was
conducted with a pair of spheres oscillating in phase.

\subsection{\label{sec:1sphere}Flow field around a single oscillating sphere}
\begin{figure}[!tb]
\begin{center}
\includegraphics[width=3in]{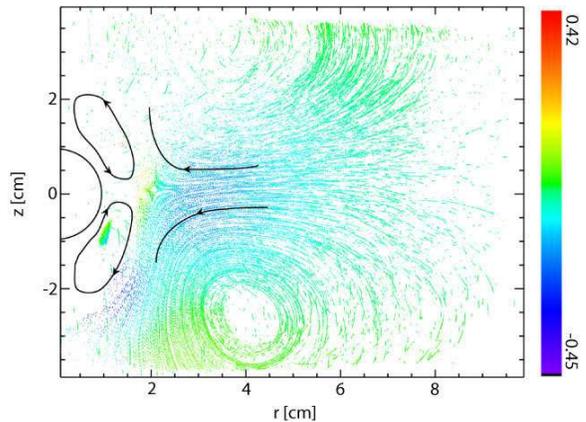}
\caption{Streaming flow pattern in the center of the tank at
$Re=30$ and $A/a=1.4$. The plot shows height vs. radial position
and the color indicates the radial velocity component, $v_r$. The
position and size of the sphere are indicated by the semicircle.}
\label{fig:model-flow}
\end{center}
\end{figure}

An example of the streaming flow around an oscillating sphere in
the center of the tank is shown in Fig.~\ref{fig:model-flow}.
Images are taken once per oscillation at the center of the stroke,
and then three dimensional positions of the tracer particles are determined. The
cylindrical symmetry of the problem allows us to present the
entire flow field by combining particles at all azimuthal angles,
$\varphi$, and plotting their height, $z$, vs. the distance from
the oscillation axis, $r$. Plots of $v_\varphi$ and horizontal
cuts confirm that the motion is almost entirely in the $r-z$ plane
in our experiments. (There are weak variations in the azimuthal
direction, a result of convection currents, which lead to crossing
tracks in regions of slow streaming flow in the z-r-plots.) The
radial velocity component $v_r$ is shown in color. A few
streamlines have been drawn in to indicate the direction of flow.
The particle tracking was optimized to observe the outer
recirculations which cause the attractive interactions between
particles. These flows are much weaker than the inner
recirculations, so this data does not resolve the inner
recirculation zones. There are some mismatched particles resulting
in a background of short spurious tracks, for example those inside
the sphere.

\subsubsection{\label{sec:velocity}Streaming velocity in the equatorial plane}
\begin{figure}[!tb]
\begin{center}
\includegraphics[width=3in]{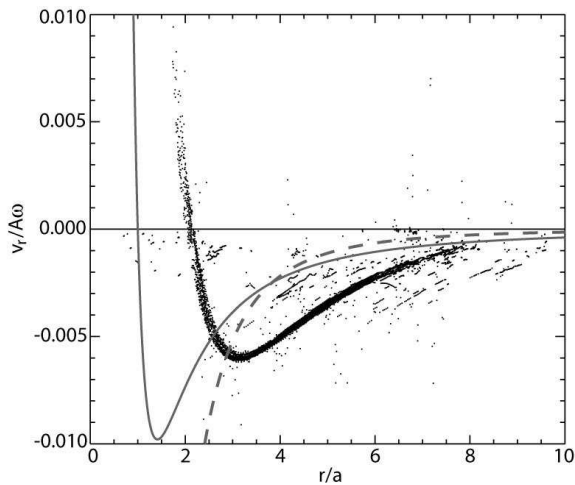}
\caption{Steady inflow velocity $v_r$ vs. r in a small stripe
around $z=0$. The scattered points are experimental data ($Re=30$,
$\epsilon=1.4$, $R_s=42$, $|M|^2=21.4$), and are compared to the
theoretical calculations of Brenner/Stone (dashed line) and Riley
(solid line; NOTE: divided by 50 to fit on the plot). Background
points are bad stereoscopic matches and should be ignored.}
\label{fig:vrvsTheory}
\end{center}
\end{figure}

Figure~\ref{fig:vrvsTheory} shows the measured radial streaming
velocity as a function of radius for particles within 0.5$cm$ of
the equatorial plane. The flow is attractive at large distances
with a stagnation point 2$cm$ from the center of the sphere. This
figure also shows two theoretical calculations of the streaming
velocity. Riley~\cite{riley:1966} predicts that flows in limit II
(from Fig.~\ref{fig:Re-eps}) have
\begin{equation}
\label{eq:riley}
\frac{u_r(r,z=0)}{A\omega}=-\frac{45}{32}\frac{\epsilon
a^2}{r^2}\left(1-\frac{a^2}{r^2}\right) \enspace,
\end{equation}
whereas Brenner and Stone~\cite{voth:2002} predict that flows in
limit III have
\begin{equation}
\label{eq:BrennerStoneInflow}
\frac{u_r(r,z=0)}{A\omega}=-0.53\sqrt{\nu/\omega}~\frac{a^2}{r^3}
\enspace.
\end{equation}
The agreement of the rate of attraction of two spheres with
Eq.~\ref{eq:BrennerStoneInflow} was a result from Voth et.
al~\cite{voth:2002}.

Figure~\ref{fig:vrvsTheory} reveals a strong discrepancy between
the theories and our experiment. The Brenner and Stone results
(dashed line) has the right order of magnitude, but a
significantly different functional form.  Riley's result (solid
line) has been divided by 50 to make it fit on the graph. The
reason for the discrepancy is partly the impossibility of
realizing an infinitesimal amplitude in the experiment.  We think
that the Brenner and Stone result does better because the
experiment at $Re_s=42$ is closer to the high $Re_s$ limit than it
is to the infinitessimal $Re_s$ limit used by Riley.  However,
viscous effects are likely still important in this $Re_s$ range.
Clearly, the inflow velocity in the range $r=a-5a$ is not
adequately captured by the available theories, and this is the
range relevant for the experiments in Voth et.
al~\cite{voth:2002}. Blackburn~\cite{blackburn:2002} has simulated
the flow around oscillating spheres and presents similar profiles
of cycle-average radial velocities.  His results also disagree
with both of the theoretical predictions, but appear to be
consistent with our experiments. However, a direct comparison is
not possible as his choices of oscillation amplitude,
$\epsilon=0.1$ and $10$, are inaccessible to the experiment.

\begin{figure}[!tb]
\begin{center}
\includegraphics[width=3in]{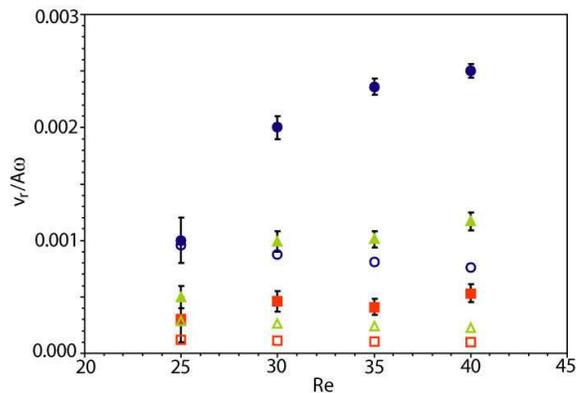}
\caption{Steady inflow velocity $v_r$ as a function of $Re$ at fixed
$\epsilon=0.7$. A comparison between the experiment (filled shapes)
and the Brenner/Stone-theory (open shapes) measured at $4a$
(circles), $6a$ (triangles), and $8a$ (squares) from the center of
the sphere.} \label{fig:vrvsRe1}
\end{center}
\end{figure}

The dependence of the equatorial velocity on $Re$ is shown in
Figure~\ref{fig:vrvsRe1}. We plot $v_r$ at $4a$ (four sphere radii
away from the center of the sphere), $6a$, and $8a$, and use
$\epsilon = 0.7$ (half that of Fig.~\ref{fig:vrvsTheory}). At each
distance, the inflow velocity increases with increasing Re. This
is in disagreement with Eq.~\ref{eq:BrennerStoneInflow} which
predicts the velocity should decrease with increasing $Re$ as
$u_r/(A\omega)=-0.53(Re/2\epsilon)^{-1/2}(a/r)^3$. This
discrepancy may be a result of viscous effects on the streaming
flow which are ignored in the potential flow solution used to
obtain Eq.~\ref{eq:BrennerStoneInflow}. Another feature of the
inflow velocity is that it does increase with increasing
oscillation amplitude $\epsilon$ at fixed $Re$~\cite{otto:2004},
but does not closely match either the linear dependence of
Eq.~\ref{eq:riley} or the square root dependence above.

\subsubsection{\label{sec:stagnationpoint}Position of the equatorial stagnation point}
\begin{figure}[!tbp]
\begin{center}
\includegraphics[width=3in]{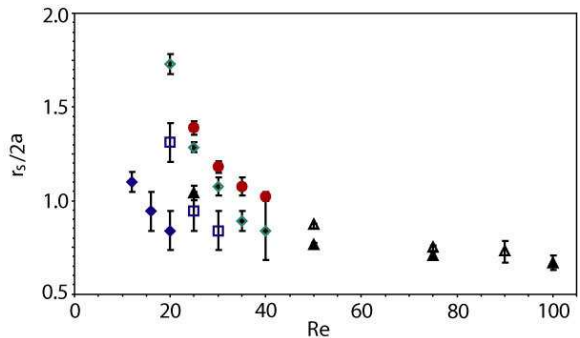}
\caption{Position of the stagnation point $r_s$ as a function of
$Re$ at different $A/a$. Filled diamond: $A/a$~=~0.16, $2a$~=~2cm;
filled triangle: $A/a=0.3$, $2a=5cm$; open triangle: $A/a=0.5$,
$2a=5cm$; filled circle: $A/a=0.7$, $2a=2cm$; open diamond:
$A/a=1.4$, $2a=2cm$; open square: $A/a=2$, $2a=2cm$. Note that
$r_s$ is measured from the center of the sphere, so the surface of
the sphere is at 0.5 on the graph.} \label{fig:rsvsRe}
\end{center}
\end{figure}

The stagnation point, where the inner and outer recirculation
zones meet in the equatorial plane, can be seen in Figures
\ref{fig:model-flow} and \ref{fig:vrvsTheory}. The position of the
stagnation point, $r_s$, as a function of the Reynolds number $Re$
is shown in Figure~\ref{fig:rsvsRe}. Several runs are shown with
different amplitudes and different sphere diameters. As $Re$ is
increasing, the stagnation point moves closer to the sphere, which
is expected since the oscillatory boundary layer thickness is
decreasing:
$\delta_{osc}=\sqrt{\nu/\omega}=a(2\epsilon)^{1/2}Re^{-1/2}$. The
position of the stagnation point already reveals a problem with
previous interpretations of the attractive interactions between
spheres. In Voth et. al~\cite{voth:2002} the spheres attract to
contact at approximately $Re<7$ and $\epsilon<1.4$. Our data, at
$Re=20$ and $\epsilon=1.4$, shows that the stagnation point is 1.7
diameters from the center of the sphere. At lower $Re$, the
stagnation point should extend farther. We conclude that the
repulsive flow inside the stagnation point extends across the
distances where attraction was observed in Ref.~\cite{voth:2002}.
Therefore, something more than simple advection by the single
particle streaming flow must be involved in the fluid mediated
attraction.

\begin{figure}[!tb]
\begin{center}
\includegraphics[width=3in]{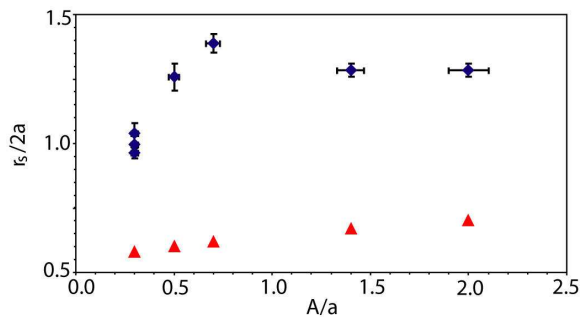}
\caption{Position of the stagnation point $r_s$ as a function of
$A/a$ at fixed $Re=25$. The triangles are one oscillatory boundary
layer thickness, $\delta_{osc}$, beyond the surface of the sphere.
The surface of the sphere is located at 0.5.} \label{fig:rsvsAa}
\end{center}
\end{figure}

The dependence of the position of the stagnation point on
oscillation amplitude is shown in Fig.~\ref{fig:rsvsAa}. The size of
the two inner recirculation zones has a maximum just below $A/a=1$,
which is in contrast to the monotonic increase in the oscillatory
boundary thickness. The reasons for this non-monotonic behavior are
not fully understood.  As the steady streaming velocity increases
with increasing $\epsilon$, the steady boundary layer may become
thinner which counteracts the increasing thickness of the
oscillatory boundary layer.  An analytic solution for the streaming
flow with non-infinitesimal oscillation amplitude would be very
helpful in further understanding of these issues.

\subsection{\label{sec:bottomplate}Changes of the flow field due to the bottom plate}
\begin{figure}[!htb]
\begin{center}
\includegraphics[width=2.3in]{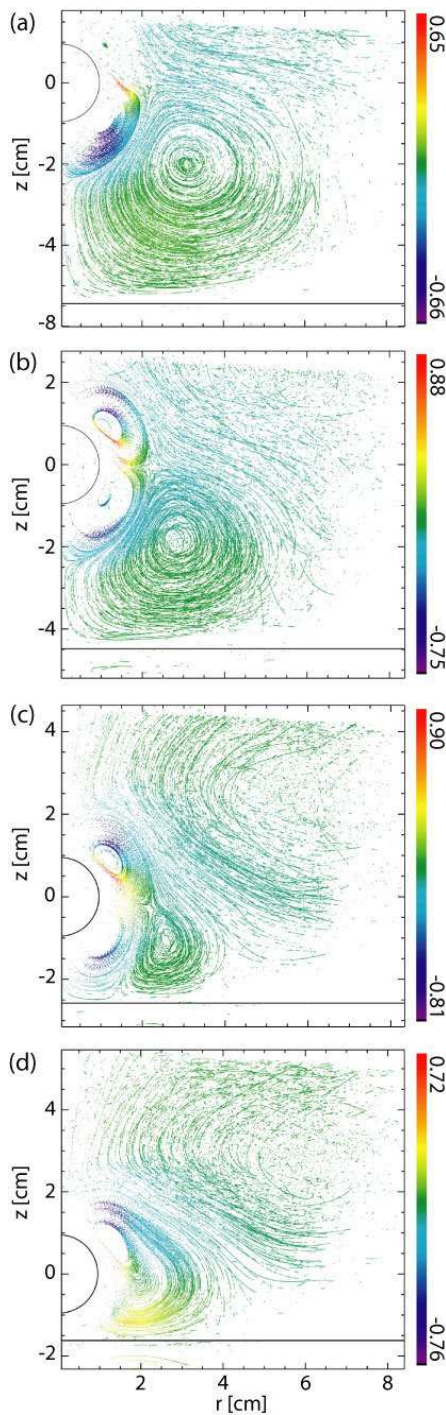}
\caption{Changes to the streaming flow geometry as the sphere
oscillates closer to the bottom wall of the container with fixed
$\epsilon=0.7$ and $Re=30$. The plot shows height vs. radial
position and the color indicates the radial velocity component,
$v_r$. For (a) $d/2a=2.0$,(b)$=1.5$, (c)$=0.5$, and (d)$=0.0$. For
each image the center of the sphere is at $z=0$, and the sphere is
at the center of the rod stroke. In (d) the sphere is almost
touching the bottom wall at the bottom of its stroke. The position
of the bottom wall is indicated by the black line.}
\label{fig:bottom-approach}
\end{center}
\end{figure}

The next set of experiments brought the sphere closer to the
bottom plate of the tank. We took 5 single sequences each having a
different distance, $d$, from the bottom plate (see
Fig.~\ref{fig:tankview}) with fixed Reynolds number and amplitude
($Re=30$, $\epsilon=0.7$). Figure~\ref{fig:bottom-approach} shows
that not only was the flow geometry distorted by the presence of
the boundary, but the topology also completely changed. The lower
of the two far field recirculation zones successively closes in,
gets compressed and eventually joins the upper of the two inner
recirculation zones. Earlier we have used the position of the
stagnation point to separate the inner repulsive recirculation and
the outer attractive flow. Here, this stagnation point has
disappeared.  An animation of the phase dependence of the flow
topology for Fig.~{\ref{fig:bottom-approach}(d) is available
online~\cite{animations}.

\subsection{Implications for interpreting fluid mediated interactions between
oscillating particles}

The streaming flow in Fig.~\ref{fig:bottom-approach} gives a
vantage point from which we can evaluate the role of streaming
flows in the fluid mediated particle interactions reported in
References~\cite{voth:2002,gollub:2004}. Because of the topology
change that unites one of the outer recirculations with an inner
recirculation, the repulsive part of the the flow extends much
farther from the sphere than it does when the wall is not present.
Even without the wall, the repulsive flow extended too far from
the sphere to explain the attraction observed in Voth et.
al~\cite{voth:2002}. Now near the bottom wall, the repulsive
region extends to $r=4a$ and covers most of the attractive range.
The current data is at higher Reynolds number ($Re=30$ compared
with $Re~10$), but Fig.~\ref{fig:rsvsRe} shows that the extent of
the repulsive recirculation extends even farther at lower $Re$.
The failure of Eq.~\ref{eq:BrennerStoneInflow} to describe the
attractive flow velocity further indicates that the attraction and
repulsion between vertically vibrated particles cannot be
explained purely by advecting the particles in the single particle
flow fields. We conclude that the interactions previously
observed~\cite{voth:2002,gollub:2004} are the result of nonlinear
interactions between the flows around multiple particles.  This
conclusion has also been drawn from the study of the horizontally
vibrated system by Klotsa et. al~\cite{klotsa:2007}.

\subsection{\label{sec:2spheres}Flow field around 2 spheres}
\begin{figure}[!tb]
\begin{center}
\includegraphics[width=3.1in]{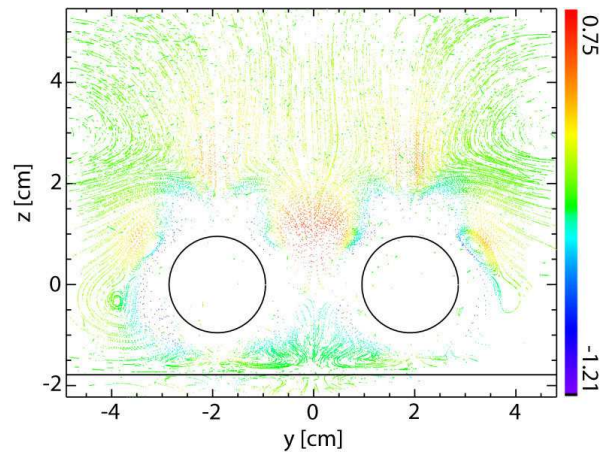}
\caption{A periodic displacement plot in a 1-cm thick layer with
$Re=30$ and $A/a=0.7$. This $x=0$-layer  is $-0.5<x<0.5$. The plot
shows $y$ vs $z$ and the color indicates the height velocity
component, $v_z$. The position and size of the spheres are
indicated by the semicircles.} \label{fig:2spheres}
\end{center}
\end{figure}

Since non-linear interactions between the flows around two spheres
must be responsible for the attractive and repulsive interactions,
we have measured the flow around two oscillating spheres. The flow
is now fully three dimensional, which makes full use of the
capabilities of 3DPTV.  We mounted two spheres on separate
vertical rods and vibrated them in phase with a fixed distance of
roughly 1 sphere diameter, $2a=1.91cm$, between them.
Figure~\ref{fig:2spheres} shows the flow around two oscillating
spheres near the bottom of the tank. Since this flow is not
azimuthally symmetric we plot only a slice of the velocity field
that contains the vertical plane going through the centers of both
spheres. Producing this image required 10 times more data than the
single particle flow fields since it could not be azimuthally
averaged. The flow geometry on the left and right edges of
Fig.~\ref{fig:2spheres} shows roughly the same topology as the
case of one sphere near the bottom wall (see
Fig.~\ref{fig:bottom-approach}). Observation of the full flow
field reveals that particles in the equatorial recirculation zone
experience azimuthal drift toward the center where they are
ejected in the upward jet. A 3D visualization of some of these
trajectories is available online~\cite{animations}.

\section{\label{sec:conclusions}Conclusions}

We have demonstrated that a simple implementation of three
dimensional particle tracking can provide accurate measurements of
the flow fields around oscillating spheres. We set out to deepen
our understanding of the fluid mediated attraction and repulsion
between particles oscillating in a fluid, building on work by Voth
et al~\cite{voth:2002}. This earlier work had observed clustering
of particles oscillated in a fluid, and found that for some
parameters there could be repulsion and intriguing dynamic states.
It also proposed that the attraction could be described by an
analytical calculation of the streaming flow around single
spheres.

For parameters typical of the experimental observations of fluid
mediated interactions, our measured velocity fields are quite
different from available analytical calculations. This discrepancy
is probably due to the theories being derived for limiting cases
that do not adequately match experimental parameters. When a
sphere is oscillating perpendicular to a flat plate, the streaming
flow undergoes a dramatic change in topology due to the
coalescence of two recirculation zones. This expands the range of
the repulsive streaming flow near the sphere. Together these
conclusions demonstrate that previous
explanations~\cite{voth:2002} gave oversimplified descriptions of
the flows that lead to the interactions between spheres.  Some
parts of the interactions can be qualitatively understood by
single particle streaming flows. However, two particle effects are
important in controlling the interactions of particles within a
few diameters of each other at the studied parameters. We have
presented measurements of the streaming flow around a pair of
oscillating spheres, but more work remains to be done in
understanding how the non-linear interaction of the oscillatory
flows leads to observed interactions between oscillating spheres.

\section{\label{sec:acknowledgements}Acknowledgements}

This work was supported by Wesleyan University and NSF Grant
DMR-0547712.  We appreciate discussions with Michael Brenner and
Jerry Gollub.


\end{document}